\documentclass[5p,twocolumn]{elsarticle}
\def\equ#1{~(\ref{#1})}
\usepackage{amsmath,amssymb}

\newcommand{\Romannumeral}[1]{\uppercase\expandafter{\romannumeral#1}}
\newcommand{\ignore}[1]{}

\allowdisplaybreaks[4]

\journal{Physics Letters B}

\begin{document}

\begin{frontmatter}

\title{Antisymmetric Tensor Fields in Randall Sundrum Thick
Branes}

\author[a]{G. Alencar \corref{cSO}}
\ead{geovamaciel@gmail.com}
\author[b]{R. R. Landim}
\author[a]{M. O. Tahim}
\author[c]{C. R. Muniz}
\author[b]{R. N. Costa Filho}
\address[a]{Universidade
Estadual do Cear\'a, Faculdade de Educa\c c\~ao, Ci\^encias e Letras do Sert\~ao Central,
Quixad\'a,Cear\'a, Brazil.}
\address[b]{Departamento de F\'{\i}sica, Universidade Federal do Cear\'{a},
Caixa Postal 6030, Campus do Pici, 60455-760, Fortaleza, Cear\'{a}, Brazil.}
\address[c]{Universidade Estadual do Cear\'a, Faculdade de Educa\c c\~ao, Ci\^encias e Letras de Iguatu, Rua
Deocleciano Lima Verde, s/n Iguatu,Cear\'a, Brazil.}

\cortext[cSO]{Corresponding author.}

\begin{abstract}
In this article we study the issue of localization of the three-form field in a Randall-Sundrum-like scenario.
We simulate our membrane by kinks embedded in D=5, describing the usual case (not deformed) and new models coming
from a specific deformation procedure. The gravitational background regarded includes the dilaton contribution. We show
that we can only localize the zero-mode of this field for a specific range of the dilaton coupling, even in the deformed case.
A study about resonances is presented. We use a numerical approach for calculations of the transmission coefficients
associated to the quantum mechanical problem. This gives a clear description of the physics involved in the model.
We find in this way that the appearance of resonances is strongly dependent on the coupling constant. We study the
cases $p=1,3$ and $5$ for $\alpha=-1.75$ and $\alpha=-20$. The first value of $\alpha$ give us one resonance peak for  $p=1$ 
and no resonances for $p=3,5$. The second value of $\alpha$ give us a very rich structure of resonances, with number deppending on
the value of $p$.
\end{abstract}


\end{frontmatter}

\section{Introduction}

From a mathematical viewpoint, anti-symmetrical tensors are natural objects living in differential manifolds. They play an important
role in the construction of the manifold's volume and therefore its orientation. The dimension of the manifold defines the sort of
possible anti-symmetrical tensors and this is used to define a specific space only of forms \cite{Nakahara:2003nw}. Besides this,
they are related to the linking number of higher dimensional knots \cite{Oda:1989tq}.

From the physical viewpoint,they are of great interest because they may have the status of fields describing particles other than the
usual ones. As an example we can cite the space-time torsion \cite{Mukhopadhyaya:2004cc} and the axion
field \cite{Arvanitaki:2009fg,Svrcek:2006yi} that have separated descriptions by the two-form. Besides this, String Theory shows the
naturalness of higher rank tensor fields in its spectrum \cite{Polchinski:1998rq,Polchinski:1998rr}. Other applications of these kind
of fields have been made showing its relation with the AdS/CFT conjecture \cite{Germani:2004jf}.

Topological defects soliton-like are studied with increasing interest in physics, not only in Condensed Matter, 
as in Particle Physics and Cosmology. In brane models, they are used as mechanism of fields localization , avoiding the 
appearance of the troublesome infinities. Several kinds of these defects in brane scenarios are considered in 
the literature \cite{yves a, yves b, yves c}. In these papers, the authors consider brane world models where the brane is supported 
by a soliton solution to the baby Skyrme model or by topological defects available in some models. As an example, in a recent paper, a model 
is considered  for coupling fermions to brane and/or antibrane modeled by a kink antikink system \cite{yves d}.

The subject of this article is to study aspects of the three-form field in a scenario of extra dimensions. Specifically we will study the issue of localization using solitons kink-like and the possibility of resonant KK modes in a Randall-Sundrum-like model (a model with thick branes) \cite{Kehagias:2000au}. The core idea of extra dimensional models is to consider the four-dimensional universe as a hyper-surface embedded
in a multidimensional manifold. The appeal of such models is the determination of scenarios where membranes have the best chances to mimic
the standard model's characteristics. In particular, the standard model presents interesting topics to study such as the hierarchy
problem, and the cosmological constant problem that can be treated by the above-mentioned scenarios. For example, the Randall-Sundrum
model \cite{Randall:1999vf,Randall:1999ee} provides a possible solution to the hierarchy problem and show how gravity is trapped to a
membrane.

In mathematical terms, the presence of one more extra dimension ($D=5$) provides the existence of many antisymmetric fields, namely
the two, three, four and five forms. However, the only relevant ones for our brane are the two and the three form. This is due to the
fact that when the number of dimensions increase, also increases the number of gauge freedom. This can be used to cancel the dynamics
of the field in the visible brane. The mass spectrum of the two and three form have been studied, for example, in Refs.
\cite{Mukhopadhyaya:2004cc} and \cite{Mukhopadhyaya:2007jn}. Posteriorly, the coupling between the two and three forms with the dilaton
was studied, in different contexts, in \cite{DeRisi:2007dn,Mukhopadhyaya:2009gp,Alencar:2010mi}. The study of this kind of coupling,
inspired in string theory, is important in order to produce a process that, in principle, could be seen in LHC. This is a Drell-Yang
process in which a pair quark-antiquark can give rise to a three(two)-form field, mediated by a dilaton. In the present case, the
coupling is needed for a different reason. As we will see in this work, the three form field without a dilaton do not has its massless
mode localized. When we introduce a dilaton, this mode can be localized and therefore this is a more interesting case to be considered.
The localization of fields in a framework that consider the brane as a kink has been studied for example in
\cite{Bazeia:2008zx,Bazeia:2007nd,Bazeia:2004yw,Bazeia:2003aw}.

This work is organized as follows. The second section is devoted to the study of the membrane as a kink. A solution of Einstein
equation is found for two scalar fields, one of them describing the kink solution and the other representing the dilaton. In the third
section we analyze the localization of the three form field with and without the dilaton, and observe that the localization is only
possible in the framework with a dilaton coupling under a specific condition. In the next section we study the gravitational background
when a deformation procedure is carried out in our original framework. We show that the massless mode of the three form is also localized
only if we regard the dilaton coupling with the same condition cited above. In the fifth section we make some considerations about
the massive modes and the possibility of resonances. Finally, in the last section, we discuss the conclusions and perspectives.

\section{The Kink as a membrane}

We start our analysis by studying the space-time background. We go
right to the point and give an explicit example through the localization of
a zero mode of the three-form mater gauge field in a four-dimensional
thick membrane embedded in five dimensions (a Randall-Sundrum like
scenario). It is well known that vector gauge fields in these kind of
scenarios are not localizable: in four dimensions the gauge vector field
theory is conformal and all information coming from warp factors drops out
necessarily rendering a non-normalizable four dimensional effective action.
However, in the work of Kehagias and Tamvakis \cite{Kehagias:2000au}, it is shown
that the coupling between the dilaton and the vector gauge field produces
localization of the later. In analogy with the work of Kehagias and Tamvakis
we introduce here the coupling between the dilaton and the
three-form field. As we comment at the end of the next section, this coupling is also necessary to localize
the three form field. Before analyzing this coupling, it is necessary to obtain
a solution of the equations of motion for the gravitational field in the
background of the dilaton and the membrane. For such, we introduce the
following action \cite{Kehagias:2000au}:
\begin{equation}
S=\int d^{5}x \sqrt{-G}[2M^{3}R-\frac{1}{2}(\partial\phi)^{2}-\frac{1}{2}
(\partial\pi)^{2}-V(\phi,\pi)].
\end{equation}
Note again that we are working with a model containing two real scalar
fields. The field $\phi$ plays the role of to generate the membrane of the
model while the field $\pi$ represents the dilaton. The potential function
now depends on both scalar fields. It is assumed the following ansatz for
the space-time metric:
\begin{equation}
ds^{2}=e^{2A(y)}\eta_{\mu\nu}dx^{\mu}dx^{\nu}+e^{2B(y)}dy^{2}.
\end{equation}
The equations of motion are given by
\begin{equation}
\frac{1}{2}(\phi^{\prime})^{2}+\frac{1}{2}(\pi^{\prime})^{2}-e^{2B(y)}V(
\phi,\pi)=24M^{3}(A^{\prime})^{2},
\end{equation}
\begin{eqnarray}
&&\frac{1}{2}(\phi^{\prime})^{2}+\frac{1}{2}(\pi^{\prime})^{2}+e^{2B(y)}V(
\phi,\pi)=\nonumber \\
&&-12M^{3}A^{\prime\prime}-24M^{3}(A^{\prime})^{2}+12M^{3}A^{
\prime}B^{\prime},
\end{eqnarray}
\begin{equation}
\phi^{\prime\prime}+(4A^{\prime}-B^{\prime})\phi^{\prime}=\partial_{\phi}V,
\end{equation}
and
\begin{equation}
\pi^{\prime\prime}+(4A^{\prime}-B^{\prime})\pi^{\prime}=\partial_{\pi}V.
\end{equation}

In order to solve that system, we use the so-called super-potential function $
W(\phi)$, defined by $\phi^{\prime}=\frac{\partial W}{\partial\phi}$,
following the approach of Kehagias and Tamvakis \cite{Kehagias:2000au}. The
particular solution regarded follows from choosing the potential $V(\phi,\pi)
$ and super-potential $W(\phi)$ as
\begin{equation}
V=\exp{(\frac{\pi}{\sqrt{12M^{3}}})}\{\frac{1}{2}(\frac{\partial W}{%
\partial\phi})^{2}-\frac{5}{32M^{2}}W(\phi)^{2}\},
\end{equation}
and
\begin{equation}
W(\phi)=va\phi(1-\frac{\phi^{2}}{3v^{2}}).
\end{equation}
In this way it is easy to obtain differential equations of first order
whose
solutions are solutions of the equations of motion above, namely
\begin{equation}
\pi=-\sqrt{3M^{3}}A, \label{pi}
\end{equation}
\begin{equation}
B=\frac{A}{4}=-\frac{\pi}{4\sqrt{3M^{3}}}, \label{B}
\end{equation}
and
\begin{equation}
A^{\prime}=-\frac{W}{12M^{3}}. \label{A}
\end{equation}
The solutions for these set of equations are the following:
\begin{equation}
\phi(y)=v\tanh(ay),  \label{dilat1}
\end{equation}
\begin{equation}
A(y)=-\frac{v^2}{72M^3}\left(4\ln\cosh(ay) +\tanh^2(ay)\right) \label{dilat2}
\end{equation}
and
\begin{equation}
\pi(y)=\frac{v^2}{4\sqrt{3M^3}}\left(4\ln\cosh(ay) +\tanh^2(ay)\right).  \label{dilat3}
\end{equation}
Following the argumentation in Ref. \cite{Kehagias:2000au}, it is possible
to see that, by the linearization of the geometry described in this section,
this model supports a massless zero mode of the gravitational field
localized on the membrane, even in the dilaton background.

Now the dilaton contribution makes the space-time singular. However this
kind of singularity is very common in D-brane solutions in string theory
(the dilaton solution is divergent). The Ricci scalar for this new geometry
is now given by
\begin{equation}
R=-\left(8A^{\prime \prime }+18A^{\prime 2}\right)e^{\frac{\pi}{2\sqrt{3M^{3}}}}
\end{equation}
where the dilaton has an important contribution. What is interesting here is
that this singularity disappears if we lift the metric solution to $D=6$. In
this case the dilaton represents the radius of the sixth dimension \cite{Kehagias:2000au}.

\section{Antisymmetric Tensor Fields and the Dilaton Coupling}

In this section we study antisymmetric tensor fields in the gravitational
background with the dilaton field. We must
look for localization of these fields in this framework. As the number of
antisymmetric tensor fields increase with dimensions, these fields should be
considered. In fact these fields have been taken in account in the
literature. The fact is that, when the number of dimension
increases, the number of gauge freedom also increases, and this can be used
to cancel the degrees of freedom in the visible brane. Therefore the only
antisymmetric tensors relevant to the visible brane are that of
rank two and three\cite{Mukhopadhyaya:2007jn}. We must focus here in the second case.
We must specifically study the string inspired coupling of the three form
field to the dilaton. As commented before this will give rise to the
localization of the field. As commented in the introduction, this kind of
coupling can give an interaction that in principle could be seen at LHC.
Despite this, we focus here on the localization properties of this framework.

We have for the action

\begin{equation}
S=\int d^{5}x\sqrt{-G}[-e^{-\lambda \pi }2Y_{MNLP}Y^{MNLP}]
\end{equation}
and $Y_{MNLP}=\partial _{\lbrack M}X_{NLP]}$ is the field strength for the
three form $X$. We can use gauge freedom to fix $X_{\mu \nu y}=\partial
^{\mu }X_{\mu \nu \alpha }=0$ and we are left with the following terms

\begin{equation}
Y_{MNAB}=\partial _{\lbrack \mu }X_{\nu \alpha \beta ]}
\end{equation}

\begin{equation}
Y_{MNAB}=\partial _{\lbrack y}X_{\nu \alpha \beta }].
\end{equation}

Using the above facts we obtain for the action

\begin{eqnarray}
S_{X}&=&\int d^{4}x\int dy\{-e^{-\lambda \pi }2[e^{-4A+B}Y_{\alpha \mu \lambda
\gamma }Y^{\alpha \mu \nu \lambda }+\nonumber \\ &&4e^{-2A-B}\dot{X}_{\alpha \mu \lambda }%
\dot{X}^{\alpha \mu \lambda }]\},
\end{eqnarray}
with  the equations of motion

\begin{equation}
\partial _{\alpha }Y^{\alpha \mu \lambda \gamma }+e^{\lambda \pi
+4A-B}\partial _{y}[e^{-\lambda \pi -2A-B}\dot{X}^{\alpha \mu \lambda }]=0.
\end{equation}

We can now separate the variables with the following ansatz

\begin{equation}
X^{\mu \nu \alpha }\left( x^{\alpha },y\right) =B^{\mu \nu \alpha }\left(
x^{\alpha }\right) U\left( y\right) =B^{\mu \nu \alpha }\left( 0\right)
e^{ip_{\alpha }x^{\alpha }}U\left( y\right),
\end{equation}%
where $p^{2}=-m^{2}$. We now write $Y^{\alpha \mu \lambda \gamma }=\tilde{Y}
^{\alpha \mu \lambda \gamma }U$, where $\tilde{Y}$ stands for the four
dimensional field strength and we get for the EM

\begin{equation}
{U}''\left( y\right) -\left( \lambda {\pi}'+2{A}'+{B}'\right)
{U}'\left( y\right) =-m^{2}e^{2\left( B-A\right) }U\left( y\right) .\label{Udilaton}
\end{equation}

A solution is found in the case $m=0$, where we find that $U=cte$
solves the above equation. In this case, the effective action can
be found easily to give us

\begin{eqnarray}
S_{X}=\int dye^{-\lambda \pi -4A+B}U^{2}\int d^{4}x[-2\tilde{Y}_{\alpha \mu \lambda
\gamma }\tilde{Y}^{\alpha \mu \nu \lambda }].
\end{eqnarray}

Using now our solution for $A$, $B$ and $\pi $ we have that, when $\lambda
>15/4\sqrt{3M^{3}}$, the integration in the extra dimension is finite.
Therefore the massless mode of the three form field is possibly localized in
a framework with a dilaton coupling.

It is worthwhile to analyze the case without the dilaton coupling. In this case the metric is given by
\begin{equation}
ds^{2}=e^{2A(y)}\eta_{\mu\nu}dx^{\mu}dx^{\nu}+dy^{2}.
\end{equation}

The gravitational solution has been found in \cite{Kehagias:2000au} and the warp factor is the same as in the case with dilaton coupling.
Therefore following the same steps as before, we can find the equation of motion
\begin{equation}
{U}''\left( y\right) -2{A} {U}'\left(
y\right) =-m^{2}e^{-2A}U\left( y\right) .
\end{equation}

Again a direct solution is found in the case $m=0$, and $U=cte$ solves the above equation.
In this particular case, the effective action can be found easily to give us

\begin{eqnarray}
S_{X}= \int dye^{-4A}U^{2}\int
d^{4}x[-2\tilde{Y}_{\alpha \mu \lambda \gamma }\tilde{Y}^{\alpha \mu \nu
\lambda }].
\end{eqnarray}

Using now our solution for $A$, we have that the integration
in the extra dimension is not finite. Therefore we see that the dilaton coupling is really essential to the localization of the zero mode.

\section{The Dilatonic Deformed Brane}

In this section we analyze a special class of solutions by defining a
deformation of the $\lambda\phi^4$ potential \cite{Bazeia:2002xg}. It is possible to
solve the equations of motion by the super-potential method. This formalism
was initially introduced in studies about non-super-symmetric domain walls in
various dimensions by \cite{DeWolfe:1999cp,Skenderis:1999mm}. The first step
in our analysis is to find the Einstein's equations for the coupled system
of the scalar-dilaton-gravity system that composes the background
space-time. We repeat here the same action already discussed in the sections
above now with the super-potential
\begin{equation}  \label{sup}
W_p(\phi)=\frac{2p}{2p-1}\phi^{\frac{2p-1}{p}}-\frac{2p}{2p+1}\phi^{\frac{2p+1%
}{p}},
\end{equation}
where $p$ is an odd integer.

By solving $\frac{\partial W_p}{\partial \phi}=\phi^{\prime }$ we arrive at $%
\phi_p(y)=tanh^p(\frac{y}{p})$ and we see that for $p=1$ we get the usual
kink solution. For $p=3,5,7...$ we can construct the so called two-kink
solutions, describing internal structures inside the membrane \cite{Bazeia:2002xg}.
The parameter $p$ introduced in the procedure controls characteristics such
as thickness and matter energy density of the membrane \cite{Gomes:2006zm}.
We must note that the equations\equ{pi},\equ{B} and\equ{A} are left unchanged. From
the super-potential cited we get the first order equations

\begin{equation}
\pi_p=-\sqrt{3M^{3}}A_p,
\end{equation}
\begin{equation}
B_p=\frac{A_p}{4}=-\frac{\pi}{4\sqrt{3M^{3}}},
\end{equation}
and
\begin{equation}
A^{\prime}_p=-\frac{W}{12M^{3}}.
\end{equation}

From the last equation we can find $A_p(y)$ \cite{Gomes:2006zm},
\begin{eqnarray}
A_p(y)=&-&\frac{v^2}{12M^3}\frac{p}{2p+1}\tanh^{2p}\left(\frac{y}{p}\right) \nonumber \\
&-& \frac{v^2}{6M^3}\left(\frac{p^2}{2p-1}-\frac{p^2}{2p+1}\right) \label{a} \\
&\times&\biggl{\{}\ln\biggl[\cosh\left(\frac{y}{p}\right)\biggr]-
\sum_{n=1}^{p-1}\frac1{2n}\tanh^{2n}\left(\frac{y}{p}\right)\biggr{\}}
\nonumber
\end{eqnarray}
and all the steps for obtaining the equations of motion and the effective action are basically identical.
For the equation of motion we therefore obtain after a separation of variables
\begin{equation}
{U}''\left( y\right) -\left( \lambda {\pi_p}'+2{A_p}'+{B_p}'\right)
{U}'\left( y\right) =-m^{2}e^{2\left( B_p-A_p\right) }U\left( y\right) ,\label{Udilatondef}
\end{equation}
and we can see that, again, we have a trivial solution for the case $m=0$. The effective action in the present case is easily found to be
\begin{eqnarray}
S_{X}=\int dye^{-\lambda \pi -4A_p+B_p}U^{2}\int d^{4}x[-2\tilde{Y}_{\alpha \mu \lambda\gamma }\tilde{Y}^{\alpha \mu \nu \lambda }].
\end{eqnarray}

Using now our solution for $A_p$, $B_p$ and $\pi_p $ we arrive in the same situation as before and have that, when $\lambda>15/4\sqrt{3M^{3}}$ and for finite $p$, the integration in the extra dimension is finite.
Therefore the massless mode of the three form field is possibly localized in a framework with a dilatonic deformed brane.

\section{The Massive Modes and Resonances}

Now we must analyze the possibility of localization for the massive modes in both, deformed and usual cases. As the usual case
can be obtained just by putting $p=1$, we must study only the more general deformed case. The best way to analyze it is
to transform the equation (\ref{Udilatondef}) in a Schr\"odinger type. It is easy to see that an equation of the form
\begin{equation}
 \left(\frac{d^2}{dy^2}-P'(y)\frac{d}{dy}\right)U(y)=-m^2Q(y)U(y),\label{pq}
\end{equation}
can be transformed in a Schr\"odinger type
\begin{equation}
 \left(-\frac{d^2}{dz^2}+V(z)\right)\bar{U}(z)=m^2\bar{U}(z),\label{schlike}
\end{equation}
through the transformations
\begin{equation}
 \frac{dz}{dy}=f(y), \quad U(y)=\Omega(y)\bar{U}(z),
\end{equation}
with
\begin{equation}
 f(y)=\sqrt{Q(y)}, \quad \Omega(y)=\exp(P(y)/2)Q(y)^{-1/4},
\end{equation}
and
\begin{equation}
 V(z)=\left(P'(y)\Omega'(y)-\Omega''(y)\right)/\Omega f^2
\end{equation}
where the prime is a derivative with respect to $y$. Now using the equation (\ref{Udilatondef}) we obtain
\begin{eqnarray}
 V(z)&=&e^{3A_p/2}\left((\frac{\alpha^2}{4}-\frac{9}{64})A'_p(y)^2-(\frac{\alpha}{2}+\frac{3}{8})A_p''(y)\right),\nonumber \\
 \quad f(y)&=&e^{-\frac{3A_p}{4}},\quad \Omega(y)=e^{(\frac{\alpha}{2}+\frac{3}{8})A_p},
\end{eqnarray}
where $\alpha=9/4-\lambda\sqrt{3M^3}$, and we take $y$ of function of $z$ through$z(y)=\int_0^yf(\eta)d\eta$. Therefore, the eingenvalues give us the masses for which the localization is possible. It is important to note that, after the above transformations, the condition for localization becomes
\begin{equation}
\int dz \bar{U}^2=finite
\end{equation}
and this is exactly the square integrable condition of the wave function in quantum mechanics. This enforce the argument that we
have a schroedinger-like equation.

\begin{figure}[ht]
\centerline{\includegraphics[scale=0.8]{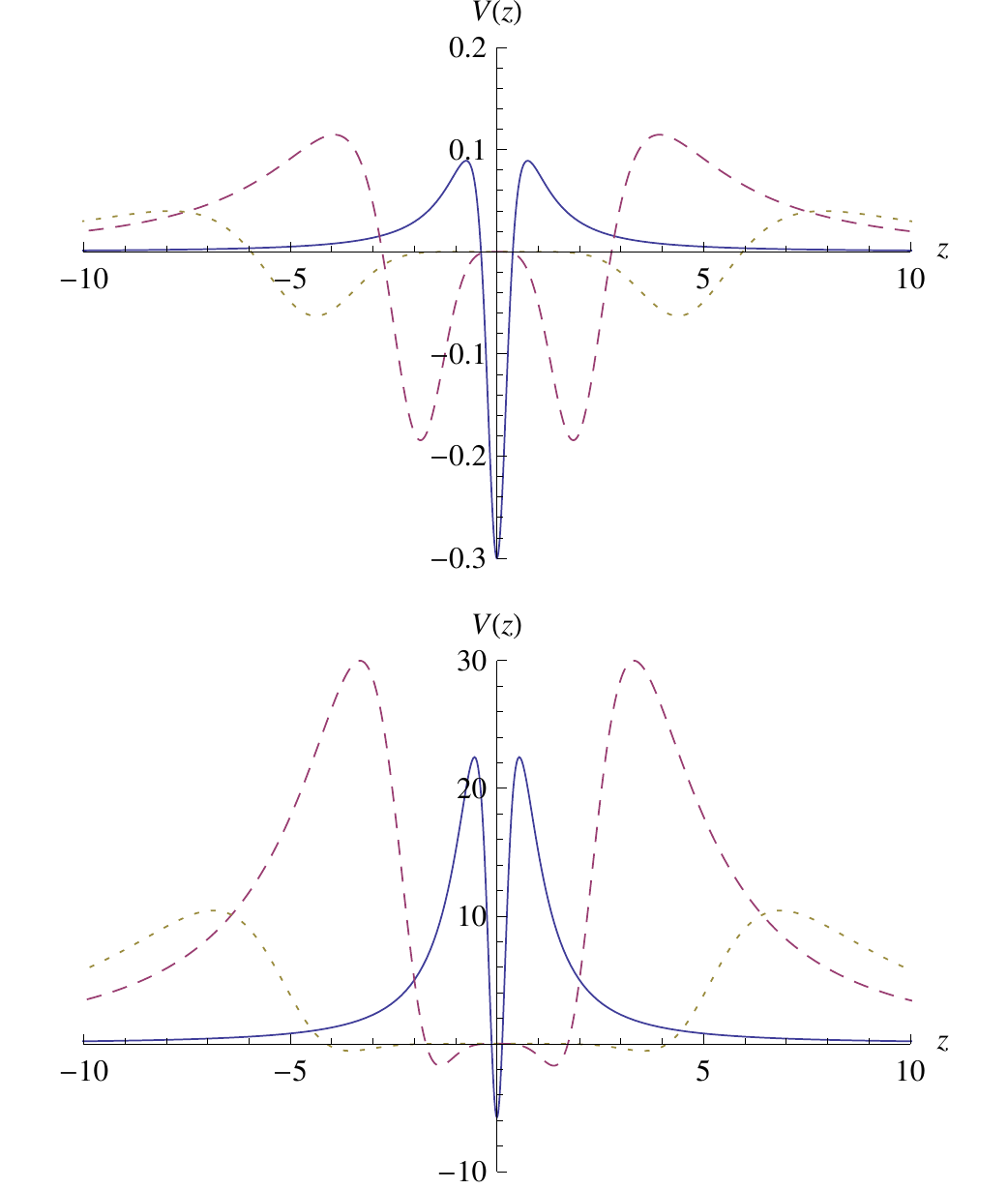}}
\caption{Potential of the Schroedinger like equation for $p=1$ (lined) scaled by 1/10, $p=3$ (dashed), $p=5$ (dotted) for  $\alpha=-1.75$ (top)
and $\alpha=-20$ (bottom).}\label{fig1}
\end{figure}

Despite its importance, the analytic solution for the above equation has not been found.
The strategy therefore is to plot numerically the potential and  analyze its shape. The appearance of resonances depends
strongly on the shape of the potential and therefore of the coupling constant. This plot
is given by fig. \ref{fig1} for $p=1,3,5$. For this we choose $\frac{v^2}{72M^3}=1$ and $v=a=1$.

Note that the shape of the potential changes significantly and this will give us a very different resonance structure. Another thing we
can observe is that the potential becomes null for $z=\infty$.Therefore
we do not have a discrete spectrum for $m>0$. Beyond this, it is a known fact from
quantum mechanics that any solution for positive $m$ must posses a oscillatory contribution, and therefore
the wave function is not normalizable. The conclusion is that the only localized mode is the
massless one. This means that this is the unique mode living inside the membrane.

Another important possibility is the appearance of resonances. As we have a volcano potential,
we can ask for the possibility that the massive modes, besides living in the extra dimension, has a peak of probability
to be found at the location of the membrane. This analysis has been done extensively in the literature \cite{Bazeia:2003aw,Cruz:2009ne}.
If we compare our Schroedinger equation with that found by these authors, we see that they are basically the same, so that
we can extract the same information about resonances of the model. The authors found that, after normalize the wave function in a
truncated region, there is a resonance very close to $m=0$, and this indicates that as lighter is the mass, bigger is the probability of
interacting with the membrane. Them we must corroborate with the analysis and the same happens in our model.

\begin{figure}[ht]
\centerline{\includegraphics[scale=0.8]{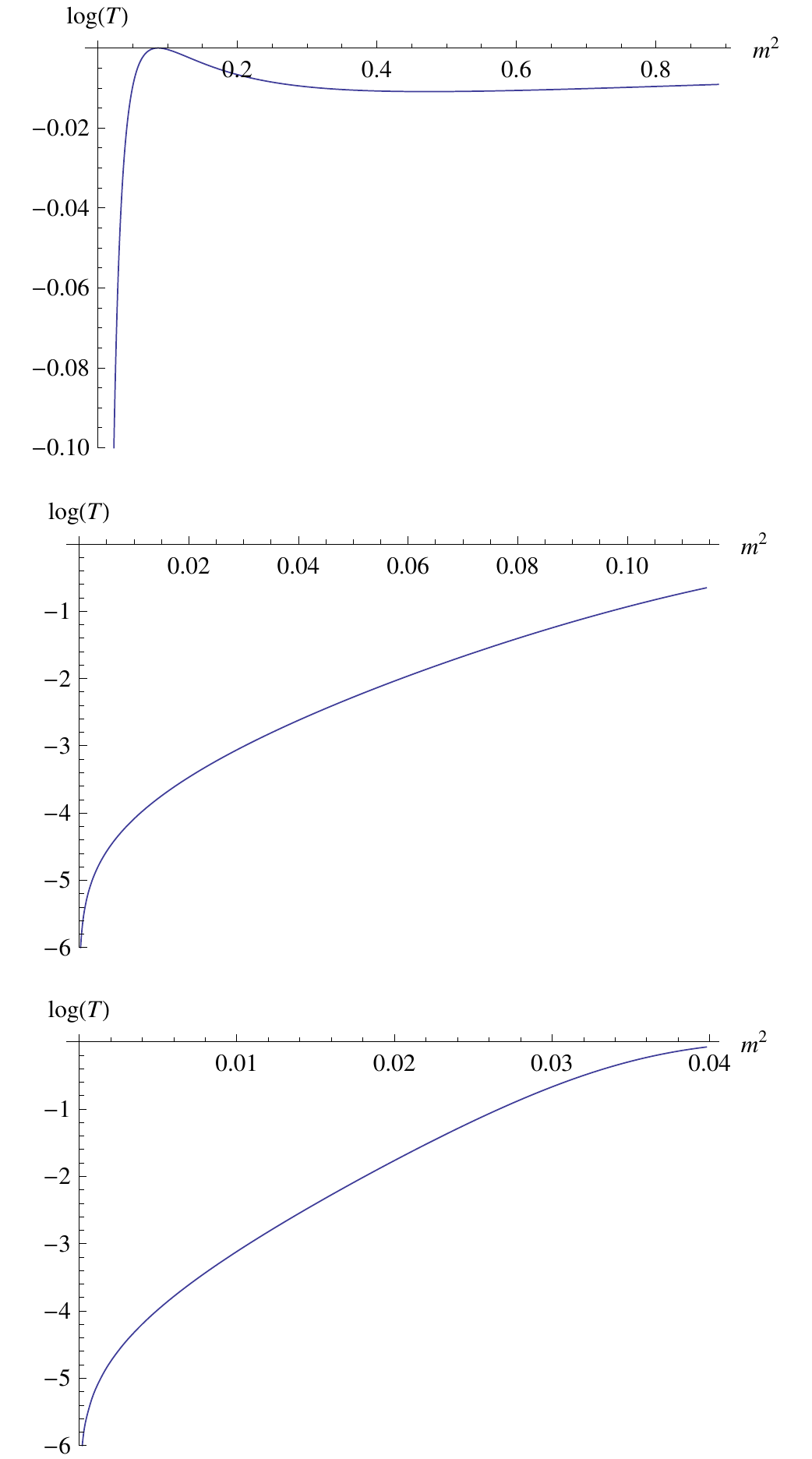}}
\caption{Logarithm of the transmission coefficient for $p=1$ (top), $p=3$ (middle), $p=5$ (bottom) for  $\alpha=-1.75$ .} \label{fig2}
\end{figure}

\begin{figure}[ht]
\centerline{\includegraphics[scale=0.8]{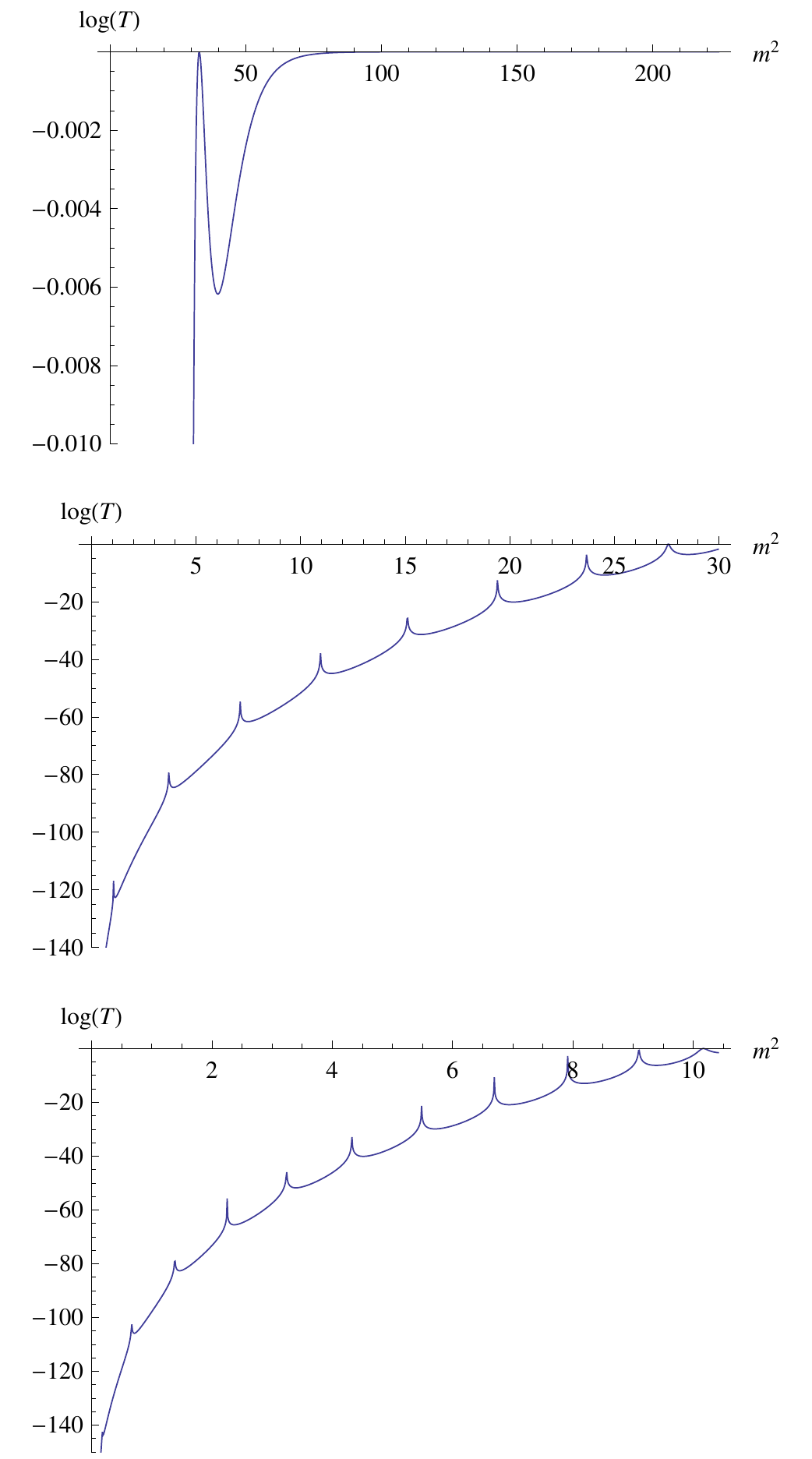}}
\caption{Logarithm of the transmission coefficient for $p=1$ (top), $p=3$ (middle), $p=5$ (bottom) for  $\alpha=-20$ .} \label{fig3}
\end{figure}

From our viewpoint the wave function has an oscillatory part and can not be normalized. To scape of this, the authors define a relative
probability, but for us it is not clear how it solves the problem. Another very important point is the truncation of the integration region.
This is equivalent to have a potential that do not falls to zero at infinity, therefore it is natural to have a peak of resonance very
close to $m=0$. This could leads to the wrong interpretation that very light modes can interact with the membrane. From our analysis, it is
not true that as lighter is the massive modes, bigger is the probability of finding them inside the membrane. In fact, this is corroborated only when $p=1$ and for the
value of the coupling constant used by the authors in \cite{Cruz:2009ne}. In this case we find, as can be seen in fig. \ref{fig2}, a resonance for the specific value $m=0.09$ . When we consider $\alpha=-20$, as shown in fig. \ref{fig3} we find a very rich resonance structure and that heavy modes can also resonate. We must point here that the Schroedinger
eq. used by them is very similar to ours but have a different multiplicative factor in the potential because they study the two form.
In principle this could change the results, therefore in a separate paper we analyze carefully that case \cite{preparation}.

In other to analyze resonances, we must compute transmission coefficients($T$), which gives a clearer and cleaner physical interpretation
about what happens to a free wave that interact with the membrane. The idea of the existence of a resonant mode is that for a given mass
the potential barrier is transparent to the particle, i. e., the transmission coefficient has a peak at this mass value. That means
the amplitude of the wave-function has a maximum value at $z=0$ and the probability to find this KK mode inside the membrane is higher.
In order to obtain these results numerically, we have developed a program to compute transmission coefficients for the given potential
profile. A more extensive and detailed analysis of resonances with transmission coefficients will be given in a separate paper by the authors \cite{preparation}.
In fig. \ref{fig2}  we give the plots of $T$ for $\alpha=-1.75,$ and considering the case $p=1,2,3$.

As said before, we find resonances for this value of $\alpha$ only for $p=1$. In fig. \ref{fig3} we show the plot for $\alpha=-20$
and we can see how this alter the existence of resonances. For this values we have an interesting structure of resonances, witch is similar to that found
in quantum mechanical problems. As we stress in our separate paper, the existence of resonances depends strongly of the shape of the potential.
In our case, therefore, the resonance is driven by the choice of the parameter $\alpha$.

\section{Conclusions and Perspectives}

In this paper we have studied the issue of localization for a three form field in a Randall-Sundrum-like model. We approach
this in a smooth space scenario, where the extra dimension is described by a kink. We show that, for the usual and the deformed
cases, we have the localization of the zero mode only for a specific range of the coupling constant given by $\lambda>15/4\sqrt{3M^{3}}$.
We study the possibility of resonances in the model. This is done throughout the numerical computations of the transmission coefficient
and we find that the appearance of peaks depends strongly on the value of the dilaton coupling constant. For high values of this constant
we find an interesting structure of resonances, very similar to what is common in quantum mechanical problems. For the value $\alpha=-20$
and $p=2$, for example, we found eight peaks of resonances. Therefore a bigger value of the coupling constant favor resonances of massive modes. This should seems natural, as the dilaton is responsible for localizing the zero mode and a bigger value of the constant would augment the possibility for existence of resonances. It should be noted that from our analysis heavy modes can also resonate.

\bigskip
We thank A. A. Moreira for providing a computer for numerical calculations. The authors
would like to acknowledge the financial support provided by Funda\c c\~ao
Cearense de Apoio ao Desenvolvimento Cient\'\i fico e Tecnol\'ogico
(FUNCAP) and the Conselho Nacional de Desenvolvimento Cient\'\i fico e Tecnol\'ogico (CNPq).

This paper is dedicated to the memory of my wife  Isa\-bel Mara (R. R. Landim)



\begin{thebibliography}{99}
%
\bibitem{Nakahara:2003nw}
  M.~Nakahara,
  ``Geometry, topology and physics,''

\bibitem{Oda:1989tq}
  I.~Oda and S.~Yahikozawa,
  ``Linking Numbers And Variational Method,''
  Phys.\ Lett.\  B {\bf 238}, 272 (1990).

\bibitem{Mukhopadhyaya:2004cc}
  B.~Mukhopadhyaya, S.~Sen, S.~Sen and S.~SenGupta,
  ``Bulk Kalb-Ramond field in Randall Sundrum scenario,''
  Phys.\ Rev.\  D {\bf 70}, 066009 (2004)
  [arXiv:hep-th/0403098]

\bibitem{Arvanitaki:2009fg}
  A.~Arvanitaki, S.~Dimopoulos, S.~Dubovsky, N.~Kaloper and J.~March-Russell,
  arXiv:0905.4720 [hep-th].

\bibitem{Svrcek:2006yi}
  P.~Svrcek and E.~Witten,
  JHEP {\bf 0606}, 051 (2006)
  [arXiv:hep-th/0605206].

\bibitem{yves a}
Y.~Brihaye, T. ~Delsate, N. ~Sawado and Y.~Kodama,
     [arXiv:hep-th/1007.0736]


\bibitem{yves b}
Y.~Brihaye, T. ~Delsate,
 Class.Quant.Grav {\bf 24},1279-1292 (2007)
   [arXiv:gr-qc/0605039]

\bibitem{yves c}
Y.~Brihaye, T. ~Delsate, B. ~Hartman
 Phys.Rev.D {\bf 74},044015(2006)
   [arXiv:hep-th/0602172]


\bibitem{yves d}
Y.~Brihaye and T.~Delsate,
 Phys.Rev.D{\bf 78}, 025014 (2008)
  [arXiv:hep-th/0803.1458]


\bibitem{Polchinski:1998rq}
  J.~Polchinski,
``String theory. Vol. 1: An introduction to the bosonic string,''
{SPIRES
entry} {\it  Cambridge, UK: Univ. Pr. (1998) 402 p}

\bibitem{Polchinski:1998rr}
  J.~Polchinski,
``String theory. Vol. 2: Superstring theory and beyond,''
{SPIRES entry} {\it  Cambridge, UK: Univ. Pr. (1998) 531 p}


\bibitem{Germani:2004jf}
  C.~Germani and A.~Kehagias,
  ``Higher-spin fields in braneworlds,''
  Nucl.\ Phys.\  B {\bf 725}, 15 (2005)
  [arXiv:hep-th/0411269].

\bibitem{Kehagias:2000au}
  A.~Kehagias and K.~Tamvakis,
  ``Localized gravitons, gauge bosons and chiral fermions in smooth spaces
  generated by a bounce,''
  Phys.\ Lett.\  B {\bf 504}, 38 (2001)
  [arXiv:hep-th/0010112].

\bibitem{Randall:1999vf}
  L.~Randall and R.~Sundrum,
  ``An alternative to compactification,''
  Phys.\ Rev.\ Lett.\  {\bf 83}, 4690 (1999)
  [arXiv:hep-th/9906064].


\bibitem{Randall:1999ee}
  L.~Randall and R.~Sundrum,
  ``A large mass hierarchy from a small extra dimension,''
  Phys.\ Rev.\ Lett.\  {\bf 83}, 3370 (1999)
  [arXiv:hep-ph/9905221].

\bibitem{Mukhopadhyaya:2007jn}
  B.~Mukhopadhyaya, S.~Sen and S.~SenGupta,
  ``Bulk antisymmetric tensor fields in a Randall-Sundrum model,''
  Phys.\ Rev.\  D {\bf 76}, 121501 (2007)
  [arXiv:0709.3428 [hep-th]].

\bibitem{DeRisi:2007dn}
  G.~De Risi,
  ``Bouncing cosmology from Kalb-Ramond Braneworld,''
  Phys.\ Rev.\  D {\bf 77}, 044030 (2008)
  [arXiv:0711.3781 [hep-th]].

\bibitem{Mukhopadhyaya:2009gp}
  B.~Mukhopadhyaya, S.~Sen and S.~SenGupta,
  ``A Randall-Sundrum scenario with bulk dilaton and torsion,''
  Phys.\ Rev.\  D {\bf 79}, 124029 (2009)
  [arXiv:0903.0722 [hep-th]].

\bibitem{Alencar:2010mi}
  G.~Alencar, M.~O.~Tahim, R.~R.~Landim, C.~R.~Muniz and R.~N.~Costa Filho,
  ``Bulk Antisymmetric tensor fields coupled to a dilaton in a Randall-Sundrum
  model,''
  arXiv:1005.1691 [hep-th].

\bibitem{Bazeia:2008zx}
  D.~Bazeia, A.~R.~Gomes, L.~Losano and R.~Menezes,
  ``Braneworld Models of Scalar Fields with Generalized Dynamics,''
  Phys.\ Lett.\  B {\bf 671}, 402 (2009)
  [arXiv:0808.1815 [hep-th]].

\bibitem{Bazeia:2007nd}
  D.~Bazeia, A.~R.~Gomes and L.~Losano,
  ``Gravity localization on thick branes: a numerical approach,''
  Int.\ J.\ Mod.\ Phys.\  A {\bf 24}, 1135 (2009)
  [arXiv:0708.3530 [hep-th]].

\bibitem{Bazeia:2004yw}
  D.~Bazeia, F.~A.~Brito and A.~R.~Gomes,
  ``Locally localized gravity and geometric transitions,''
  JHEP {\bf 0411}, 070 (2004)
  [arXiv:hep-th/0411088].

\bibitem{Bazeia:2003aw}
  D.~Bazeia, C.~Furtado and A.~R.~Gomes,
  ``Brane structure from scalar field in warped spacetime,''
  JCAP {\bf 0402}, 002 (2004)
  [arXiv:hep-th/0308034].

\bibitem{Bazeia:2002xg}
  D.~Bazeia, L.~Losano and J.~M.~C.~Malbouisson,
  ``Deformed defects,''
  Phys.\ Rev.\  D {\bf 66}, 101701 (2002)
  [arXiv:hep-th/0209027].

\bibitem{DeWolfe:1999cp}
  O.~DeWolfe, D.~Z.~Freedman, S.~S.~Gubser and A.~Karch,
  ``Modeling the fifth dimension with scalars and gravity,''
  Phys.\ Rev.\  D {\bf 62}, 046008 (2000)
  [arXiv:hep-th/9909134].



\bibitem{Skenderis:1999mm}
  K.~Skenderis and P.~K.~Townsend,
  ``Gravitational stability and renormalization-group flow,''
  Phys.\ Lett.\  B {\bf 468}, 46 (1999)
  [arXiv:hep-th/9909070].


\bibitem{Gomes:2006zm}
  A.~R.~Gomes,
  ``Gravity on the bloch brane,''
  arXiv:hep-th/0611291.

\bibitem{Cruz:2009ne}
  W.~T.~Cruz, M.~O.~Tahim and C.~A.~S.~Almeida,
  ``Results in Kalb-Ramond field localization and resonances on deformed
  branes,''
  Europhys.\ Lett.\  {\bf 88}, 41001 (2009)
  [arXiv:0912.1029 [hep-th]].

\bibitem{preparation}
G.~Alencar, R.~R.~Landim, M.~O.~Tahim, C.~R.~Muniz and R.~N.~Costa Filho,
In preparation.
\end{thebibliography}
\end{document}